# Rain Streak Removal in a Video to Improve Visibility by TAWL Algorithm

Muhammad Rafiqul Islam, *Student Member IEEE* and Manoranjan Paul, *Senior Member IEEE*

*Abstract*—**In computer vision applications, the visibility of the video content is crucial to perform analysis for better accuracy. The visibility can be affected by several atmospheric interferences in challenging weather-one of them is the appearance of rain streak. In recent time, rain streak removal achieves lots of interest to the researchers as it has some exciting applications such as autonomous car, intelligent traffic monitoring system, multimedia, etc. In this paper, we propose a novel and simple method by combining three novel extracted features focusing on temporal appearance, wide shape and relative location of the rain streak and we called it TAWL (*Temporal Appearance, Width, and Location*) method. The proposed TAWL method adaptively uses features from different resolutions and frame rates. Moreover, it progressively processes features from the up-coming frames so that it can remove rain in the real-time. The experiments have been conducted using video sequences with both real rains and synthetic rains to compare the performance of the proposed method against the relevant state-of-the-art methods. The experimental results demonstrate that the proposed method outperforms the state-of-the-art methods by removing more rain streaks while keeping other moving regions.**

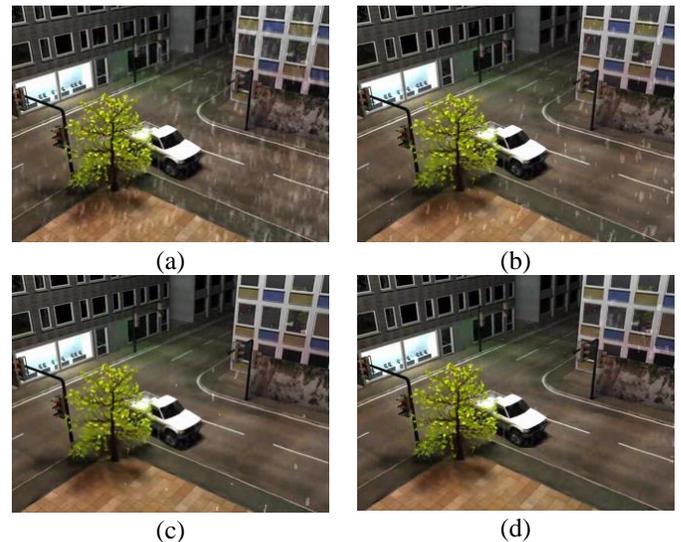

Fig. 1. Low visibility scenario due to rain and the performance of the rain removal algorithms to make rain free video where (a) a video frame with rains, (b) rain free frame using an existing algorithm [18], (c) rain free frame using another existing algorithm [19], and (d) rain free frame using the proposed TAWL method.

## I. INTRODUCTION

THE visibility of a video is affected by many atmospheric interferences to degrade the quality of the video content. The video information is also affected by a climatic catastrophe such as rain [1-6]. The low visibility situation degrades the performance of subsequent video analysis or processing applied in computer vision techniques. This undesirable situation degrades the performance of several computer vision applications such as driverless car, intelligent traffic monitoring system and surveillance [7-9]. As a result, it is a necessary task to improve the visibility of a video affected by external things like rain.

Many types of numeric methods have been proposed to improve the visibility of images/videos captured with rain streak nosiness [10-19]. They can be categorised into two classes: multiple images/video-based approaches and single-image based methods.

The scope of the paper is to remove rain streaks from video sequences. Fig.1 shows an example of the performance of the proposed TAWL method to generate a rain-free frame compared to the outcome using other existing methods.

Garg *et al*. [10] firstly raised a *Rain Streaks Removal in Video* (RSRV) method with a comprehensive analysis of the visual properties such as spatial distribution, shape and velocity of the raindrops on an imaging system. They proposed that two camera properties, exposer time and depth of field adjustment can reduce or even remove the effects of rain in a video sequence. Subsequently, many approaches have been recommended for the RSRV task and achieved a good result in rain streak removing with a variety of rain conditions. Wide-ranging primary video-based methods are reviewed in [11]. Some very active scenes have been studied in [12]. Where Kim *et al*. [13] have focused on the time-based relationship of rain streaks and the low-rank characteristic of rain-free videos. Santhaseelan *et al*. [14] marked and eliminated the rain streaks based on phase congruency features. You *et al*. [15] worked with the environments where the raindrops are situated on the window glass or the windscreen of the car. In [16], the authors focused on the directional property to propose a tensor-based RSRV method. Ren *et al*. [17] worked for both conditions, snow and rain. Authors considered matrix decomposition technique. Wei *et al*. [18] have stochastically modelled the rain streaks and have not considered the deterministic features. The rain-free background has been modelled using a mixture of Gaussians while the multi-scale convolutional filters are introduced by Li *et al*. [19] from the rainy data. Both methods reached to the satisfactory level of performances with surveillance videos.

The methods based on deep learning for the RSRV also started to reveal their effectiveness [4, 20-23]. Most of the proposed models have been developed based on single-image features but can be applied on video sequence as well. A video sequence is a combination of a single image called frame. These models have mostly addressed the interference/difficulties in visibility caused by rain streaks accumulation. In [24], authors use binary mapping representing rain and without rain pixels to train the rain model. In the rain removal method, they used a contextualised dilated network. A GAN based deraining method has been proposed in [1]. Authors have also introduced residual learning to develop a rain removal model. A two-stage RNN architecture has been proposed for video deraining by [22]. A sequential deep unrolling framework has been proposed to exploit spatial and temporal features for video deraining [23].

However, not all insightful characteristics possessed by rain streaks in a video still have not been explored. The literature explores that in video rain removal techniques, two difficulty aspects of the rain removal from a video are to distinguish rain streaks from the moving regions and remove the rain streak completely. The existing methods can perform better in one or other aspect but not at the expected level in both aspects. Some of the existing methods cannot remove rain streak in the real-time for a video as they may need to store future frames to learn/train the models. Moreover, some existing methods cannot perform at the expected level if the frame rate and resolution of the video have been changed.

The insightful characteristics such as temporal duration, relative position within a frame, shape, location, etc. of the rain streak are crucial to distinguish the rain streak from the background and moving regions for making a video free of rain streak. In this study, we propose a novel approach to remove the rain streak from a video to produce a better video by exploiting novel characteristics of the rain, such as appearance duration, shape, and location. The proposed method progressively learns the background, identifies rain streak by utilising features, and removes rain streak frame by frame. As our method extracts features for every current frame, which is being processed, this method can be applied in real-time application with the consideration of some processing time.

First, we modelled the background with *Low-Rank Matrix factorisation* (LRMF) and applied a *Mixture of Gaussian* (MoG) to separate background and foreground [19][25][33]. After separation of the background, the foreground usually includes rain streak and moving objects. The main challenge of the rain removal algorithms is to separate rain streak from the moving object for the rain-free video. Here we model the rain streak based on *temporal appearance* (TA) of the rain streak in the video. We have observed that the presence of rain in the identical location in head-to-head frames of a video sequence is highly unlikely. We have considered this property to distinct rain streak from other moving objects bearing in mind about the impact of the frame rates of the video. But it misses out some portions of the moving objects which also have a shorter appearance. To solve this issue and recover the missing moving objects, we extract more features based on the other properties of the rain streak. We exploit the width of the rain streak which filters some false positive from the candidate rain pixels. We have observed that the rain streak has a range of specific widths bearing in mind about the resolution impact. By combining TA and width properties provide a good rain removal performance, however, still some false positive is detected. The candidate pixels location is also a significant property to distinguish the rain streak from the moving objects. We also exploit location-wise properties to identify rain from other moving regions.

In this paper, we propose a novel algorithm by combining the TA properties of rain streak with the shape and location properties of rain streak to improve the recovered moving objects in rain-free videos. We also see the performance of the proposed method in different resolutions, and frame rate as the TA, shape and location properties have been changed with frame rates and/or resolutions. To make the proposed method effective in different resolutions and frame rates, we have used properties criteria in an adaptive fashion.

The preliminary idea based on the TA is published in a conference paper [7]; however, our contributions are:

- We introduced and formulated the temporal appearance (modified compared to [7] in the light of frame rate invariant) of the rain streak to differentiate them from other moving objects.
- In addition, we also developed and formulated the two other vital features of the rain streak based on the shape and location where the correlation of the neighbouring pixels of the already identified rain streak has been exploited.
- We have fused different criteria to make the final decision.
- We have adaptive thresholds to make the proposed method effective in different resolutions and frame rates.

## II. METHODS

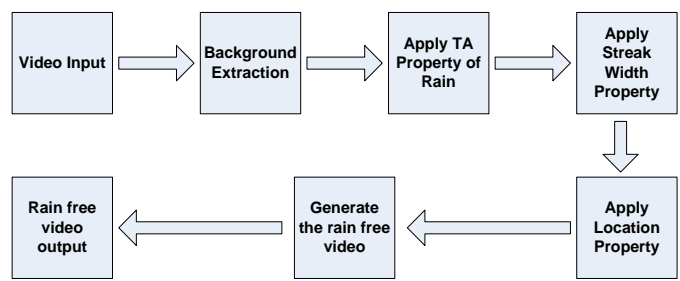

Fig. 2. Block Diagram of the proposed TAWL rain streak removal method.

The proposed rain streak removal algorithm contains five significant steps (see the schematic diagram in Fig. 2): (i)

background extraction, (ii) apply rain streak TA property, (iii) apply width property, (iv) apply location property, and (v) generate rain-free video. In this method, we explore insightful rain streak properties to refine the candidate rain streak pixels. First, we separate all moving regions, including rain and moving objects from a current frame using the generated background frame. The background frame is generated from the past frames of a video using an existing method [19][25][33]. To separate the rain streaks from other moving regions, we exploit TA property of rain streaks which may include some short-appeared moving regions with rain streaks. Then we apply width property to filter out moving regions which are short appeared but relatively bigger in size as the rain streaks usually are smaller in width. Finally, we apply the location of candidate pixels to filter out the false-positive selection of rain streak as the rain streaks are generally isolated and scattered. All the five significant steps are discussed in the following sections in details.

### A. Background Modeling

Many dynamic background modelling approaches [25-27] are available and basic concept to develop these modelling is very similar. The background remains the same over all the frames in a video scene captured by a static camera except the interference of moving objects and change of light. Thus, this background layer can be formulated as recovering a low-dimensional subspace [28-32]. The regular approach to subspace learning is the subsequent low-rank matrix factorisation (LRMF):

$$B = Fold\ (UV^T) \qquad (2)$$

where, $U \in R^{d \times r}, V \in R^{n \times r}, d = hw, r < \min(d, n)$, and the operation of '*Fold*' refers to fold up each column of a matrix into the corresponding frame matrix of a tensor.

At each frame, we generate a background frame. We use the background frame to find rain streak and other foregrounds as well to generate the rain-free video in the proposed method.

Initially, we have generated foreground by subtracting the background from the input frame,

$$F_n = \begin{cases} 1, & |I_n - B_n| > Threshold \\ 0, & otherwise \end{cases} \qquad (7)$$

where $F$ is a foreground binary image of the $n^{th}$ frame, $I_n$ is the original $n^{th}$ frame and $B_n$ is the background frame at the $n^{th}$ frame. Here we use an intensity threshold value 20 to eliminate the effect of other light or illumination interference from the generated foreground. This image contains rain streak and moving objects.

### B. Temporal Appearance (TA) with Frames Rate Invariant

After subtracting the background frame from the current frame, we can get the foreground which comprises both rain streaks and moving objects. Fig. 3(a) shows the original 85$^{th}$ frame, Fig. 3(b) shows background frame at frame 85$^{th}$ frame and Fig. 3(c) shows the rain streaks and moving objects at frame 85 in greyscale for the *Traffic* video sequence. The figure demonstrates that the background modelling with the threshold successfully detected moving regions, including rain streak. To separate the moving objects from the rain streaks, we exploit the TA property of the rain streak in a video sequence. We have observed that the appearance of the rain streak in the same location in adjacent frames of a video sequence is highly unlikely. We have exploited this property to separate rain streak from other moving objects.

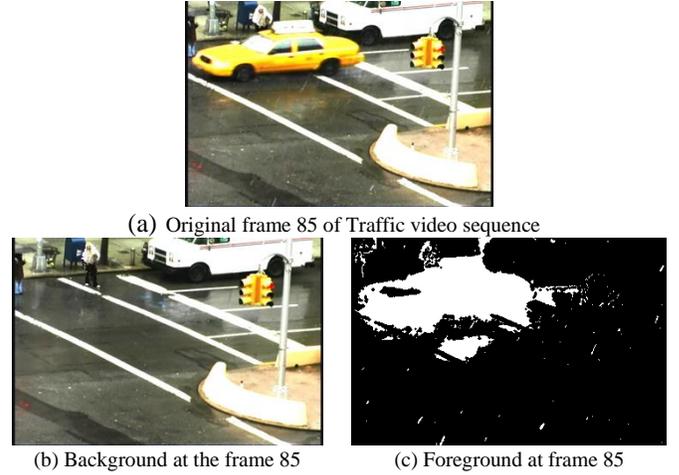

(a) Original frame 85 of Traffic video sequence

(b) Background at the frame 85     (c) Foreground at frame 85

Fig. 3 Results of background modelling [19] using the Traffic video sequence to demonstrate the separation of moving regions, including rain streak from the background using dynamic background modelling.

The red circles marked in Fig. 4 demonstrate the TA property of the rain streak. Two adjacent frames (Frame 84 and Frame 85) represent rain streaks in four locations of each frame. Rain streaks appear at two red circles (i.e. top and top right) in Frame 84 but disappear in Frame 85. Rain streaks do not appear at two red circles (i.e. bottom and left bottom) in Frame 84 but appear in Frame 85. This observation demonstrates that the rain streak appears at a particular location of a frame in a video for a very short time and may comprise few frames depending on the frame rate of the capturing devices. However, the moving object does not show the low appearance characteristic like rain streak in an area normally. Rain streak appears in a video discreetly; normally it changes location frequently for low to mid-intensity of rain. In comparison, the moving object changes the location smoothly (see the moving car).

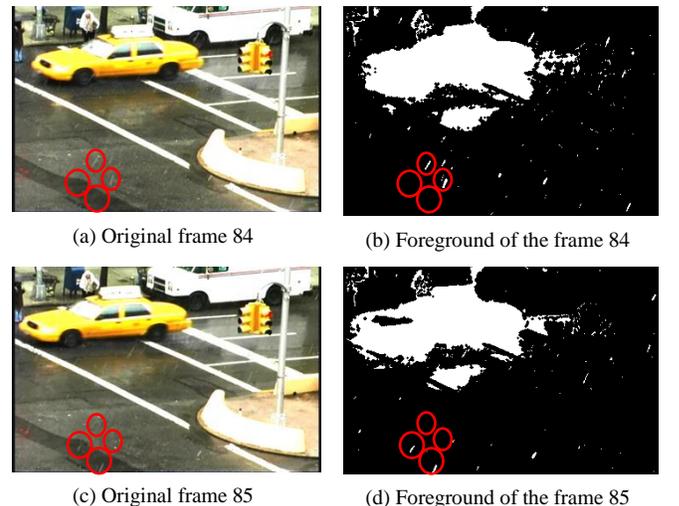

(a) Original frame 84     (b) Foreground of the frame 84

(c) Original frame 85     (d) Foreground of the frame 85

Fig. 4: Observation of temporal appearance property of rain streak in the Traffic video sequence.

We have used this temporal appearance characteristic of rain streak to separate rain streak from the moving objects of the foreground. To model rain streak and analyse the temporal feature, we have generated a mask using the binary image *F* for each frame against its adjacent previous *m* number of frames. In the binary image, '1' represents a foreground comprising rain and other moving objects, and '0' represents the background.

$$M_n = \sum_{i=n}^{n-m} F_i; (i = n, n-1, n-2, ... n-m) \quad (1)$$

where *M* represents a mask of the $n^{th}$ frame, *F* represents the foreground binary images of adjacent frames (described above), and *m* is the maximum number of adjacent frames. We use the previous *m* number of frames to make the decision contemporary as the scene may be changing significantly enough so that the mask may not be relevant to represent the recent changes. We consider every pixel location's appearance value in the mask. If the appearance value is more than a certain duration threshold in terms of the frame rate of the video, it is considered as the part of object area and any value more than zero and up to that duration threshold is considered as the rain area; otherwise, it is considered as background area. We use the duration threshold as the 25% of the frame rate to classify the rain, object, and background areas as the appearance duration vary with capturing frame rates (see explanation below). In Fig. 5, all yellow-coloured area is considered as the object area, the red coloured area as rain streak area and the blue area as a background area.

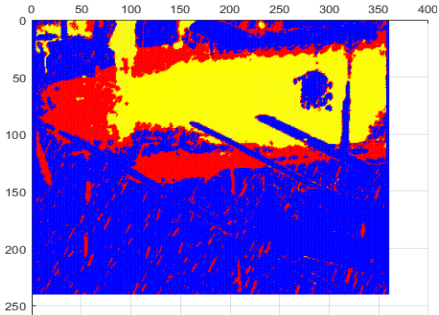

Fig. 5: Mask of 95th frame represents background, rain and moving objects with blue, red and yellow coloured areas respectively.

The TA value mostly depends on the frame rate because if the capture device is operating in a high frame rate, a rain streak may appear a greater number of frames. That's why for better rain removal, we need to make the proposed method is invariant of frame rates or in other word the proposed method should be applicability in different frame rates. Thus, the threshold we have used against the mask *M* is a function of the frame rates so that the threshold can be adaptive with the frame rate for the successful of rain removal. Fig. 6 shows the effect of different frame rates of a video if we use a constant value of the threshold in different frame rates. For the video with higher frame rate, the loss of moving object is less compared to that of the lower frame rate. The results are different for the different frame rate of the video with a fixed threshold. Thus, we can exploit the TA property successfully using an adaptive threshold in terms of different frame rates. The rain-free frame and the rain streak detected using only the TA property are shown in Fig. 7.

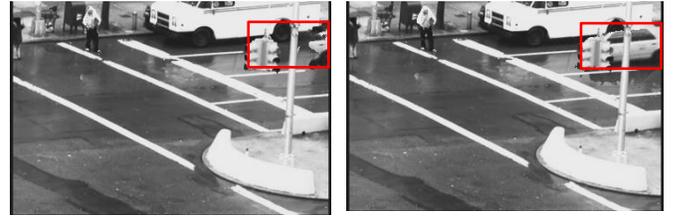

(a) At a lower frame rate  (b) At a higher frame rate

Fig. 6: The effect of the different frame rates on TA output with a constant value of threshold of TA modelling using the Traffic video sequence.

*C. Width and Location Properties for Frame Rate and Resolution Invariant*

After applying rain streak model based on the TA feature of rain streak, we missed a few portions of moving objects in rain-free video frame as the model misclassifies some parts of the object as rain streak (Fig. 7). For example, a portion of the moving car is also identified as rain streak in the TA process (see Fig.7 (b)) together with the rain streak. To overcome this issue and reduce the false positive due to the inclusion of moving objects in the outcome of the TA process, we need to filter out the moving objects from the identified rain streak using other properties of the rain streak. We refine the candidate rain streak pixels by two consecutive filters. Before applying these filters, we created two binary images for each frame based on TA properties. One consist of candidate rain streaks and others includes moving objects. We modelled both filters based on rain streaks characteristics. One rain streak's width and another one is the relative position of the candidate pixels. They are discussed in detail below:

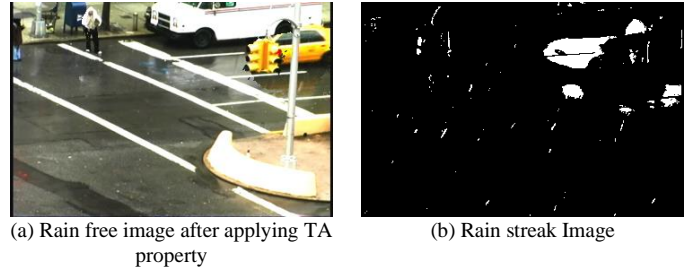

(a) Rain free image after applying TA property  (b) Rain streak Image

Fig. 7: Outcome of the rain streak model based on TA property only.

Rain Streak Width**:** From the candidate rain streaks, we filtered out the false positive by measuring the rain streak width. The rationality for using the width of the rain streak is that the width of the rain streak is not very wider and normally discrete. We check the number of consecutive '1' in every row of the binary image includes candidate rain streaks. In this filter, we consider a *length* threshold with respect to the frame width as a maximum rain streak width. Thus, any rain streak candidate with a number of consecutive '1' in a row is less than the length threshold of the frame width is considered as rain streak. We use 5% of the width of the frame as the length threshold. We experimentally observed many datasets with different types of rain streaks, and we found our consideration works better. We consider the threshold against the width of the frame to make the filter resolution-independent so that the adaptive threshold should work in different resolutions.

The pixel value of rain streak width highly depends on the image resolution. We experimentally observed the effect of different resolutions of a video sequence. Fig. 8 shows the results with different resolutions of a video sequence where the threshold of rain streak width considered as a constant value. The figure demonstrates that for higher resolution video, the rain streak width filter missed some rain streaks by considering them as objects as their pixel size is increased due to the larger resolution. On the other hand, the rain streak width filter can select more rain streaks at a lower resolution.

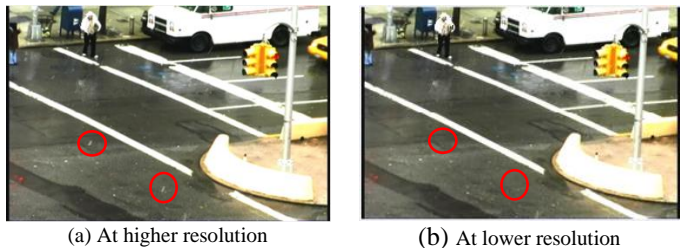

(a) At higher resolution      (b) At lower resolution
Fig. 8: The effect of different resolutions on the final results with invariant rain streak width

Rain Streak Location: After refining with the width property, some false positives are filtered out; however, still some false positives remain as it includes some moving regions with the similar width of the rain streak. For further filter out the moving areas, we check the neighbourhood pixels of the candidate rain streaks to determine how close they are to the moving objects. If they are very close and connected with the moving objects, then it is considered as a part of the moving objects rather than rain streak, thus need to be filtered out as false positive. The neighbour pixels of each candidate pixels are checked using the separated binary objects image. The same number of neighbour pixels in each quadrant (up-right, up-left, down-left and down-right) of the candidate pixel is checked to see whether they are an isolated cluster of '1's or not. If they are an isolated cluster of '1's then we assume that they are rain streak; otherwise, they are a part of a moving object. The rationality of this assumption is that if the cluster of '1's has connected with another cluster of '1's, then they are part of a moving object rather than rain streak as the rain streak is usually isolated.

### D. Rain-Free Video Generation

After applying all those extracted features, we have generated an object mask for the current processing frame. To generate a rain-free video frame, we have used both the generated background frame at the current frame position and the current frame. For example, if we like to generate the rain-free frame for the current $i^{th}$ frame, then we use both $i^{th}$ background and the $i^{th}$ frame. Through the processes as mentioned earlier, we identify each of the pixels as a background, rain and moving object. For a rain-free frame, if the pixel is identified as a background or rain, then the corresponding pixel intensity is taken from the background frame, and if the pixel is identified as the moving object, then the corresponding pixel intensity is taken from the current frame.

### III. RESULTS

We have conducted experiments using video sequences with real rain to compare the performance of the proposed method and other contemporary and relevant methods. This comparison provides a subjective quality assessment as there is no ground truth of the rain-free real videos. We also compare the performance using video sequences with synthetic rain to understand subjective and objective measurements as the synthetic video sequences have ground truth to compare with.

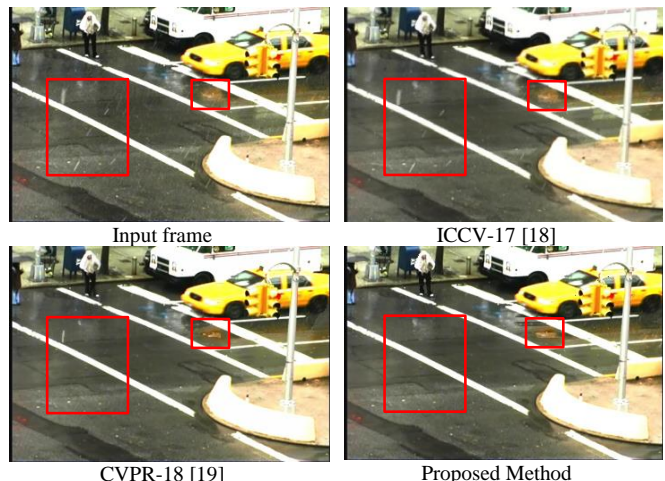

Input frame      ICCV-17 [18]

CVPR-18 [19]      Proposed Method
Fig. 9. Rain removal results and comparison between the Proposed method and other methods of video sequence "traffic".

We have used two existing methods [18, 19] to compare the performance of the proposed method. These two methods are based on the feature extractions, relatively recent and relevant to the proposed method; thus, we select these two methods to compare with. The feature extraction with physical meaning gives us a better understanding of the rain characteristics for classification.

#### A. Experiments on Real Rain Video

Fig. 9 shows the experimental results of a video sequence "traffic" at the frame 72. The figure demonstrates that the proposed method outperforms these two methods in both cases rain removal and object recovery. The identified areas using red rectangles show that some distortions have been occurred in CVPR 18 [19] method's results. This portion is a part of object reflection. Moreover, the proposed method successfully removes more rain streaks compared to other methods by keeping moving regions quality better.

Fig. 10 shows the results of a video sequence called "highway" at frame 97. The proposed method performs better than all methods. The proposed method has removed more rain streaks than the other methods. The rectangle and circle marked areas clearly show that the proposed method can remove more rains.

Fig. 11 shows the results of video sequence called "wall", at frame 15. The proposed method outperforms the other two methods. The proposed method has removed more rain streaks and generated high-quality rain-free video frame compared to other methods [18-19].

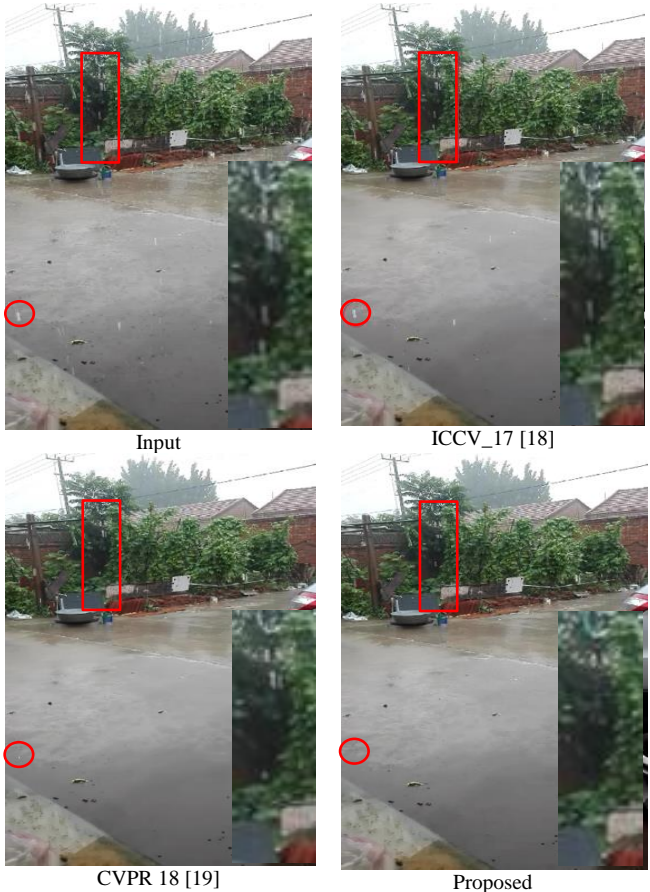

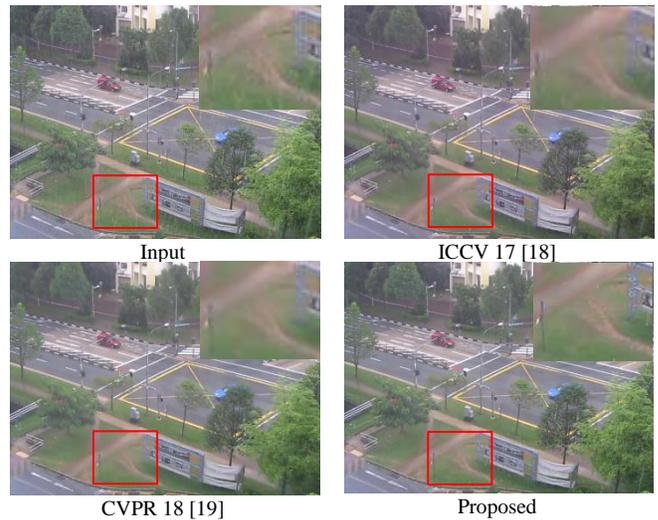

Fig. 10. Rain removal results and comparison between the proposed method and other methods of the Yard video sequence.

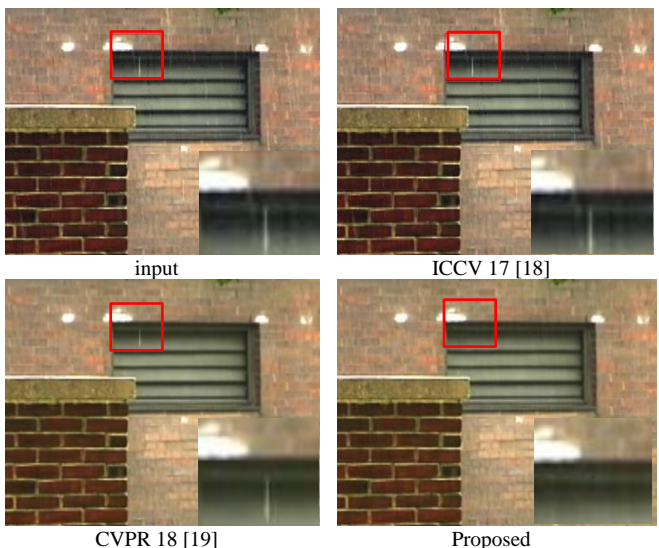

Fig. 11. Rain removal results and comparison between the proposed method and other methods for a video sequence called "wall".

Fig. 12 shows the results of a video sequence called "Ra4", at frame 35. The proposed method outperforms these two methods. The selected area shows that the rain-free frame is cleaner in the result of the proposed method.

Fig. 12. Rain removal results and comparison between the proposed method and other methods for a video sequence called "Ra4".

### B. Experimental results of Synthetic Rain streak

To understand the performance of the proposed method against the other two methods, we also provide experimental results using videos with synthetic rain. Fig. 13 shows the results of a synthetic video sequence called "truck" at frame 65. The proposed method can remove almost all rain streaks while other methods fail to remove rain streak in several areas.

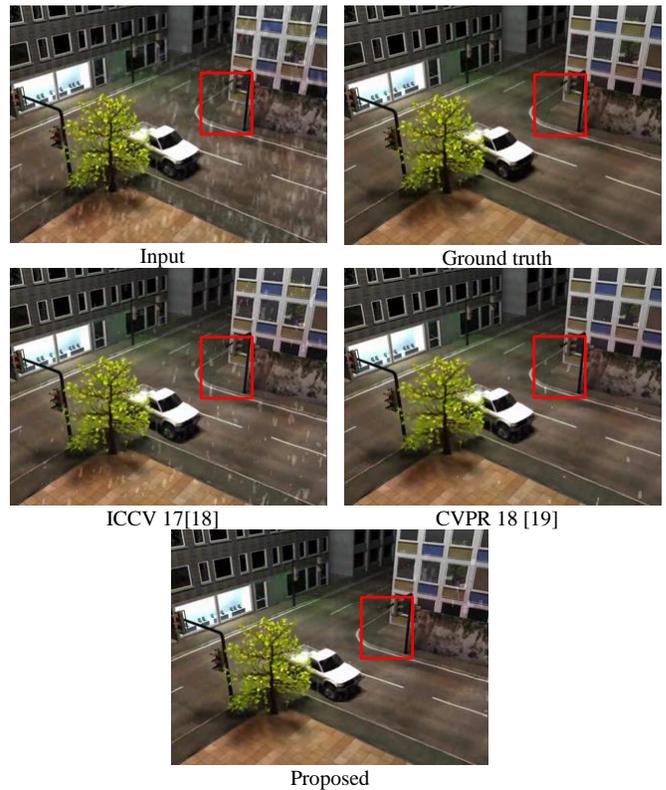

Fig. 13. Rain removal results and comparison between the Proposed method and other methods using a video sequence called "truck".

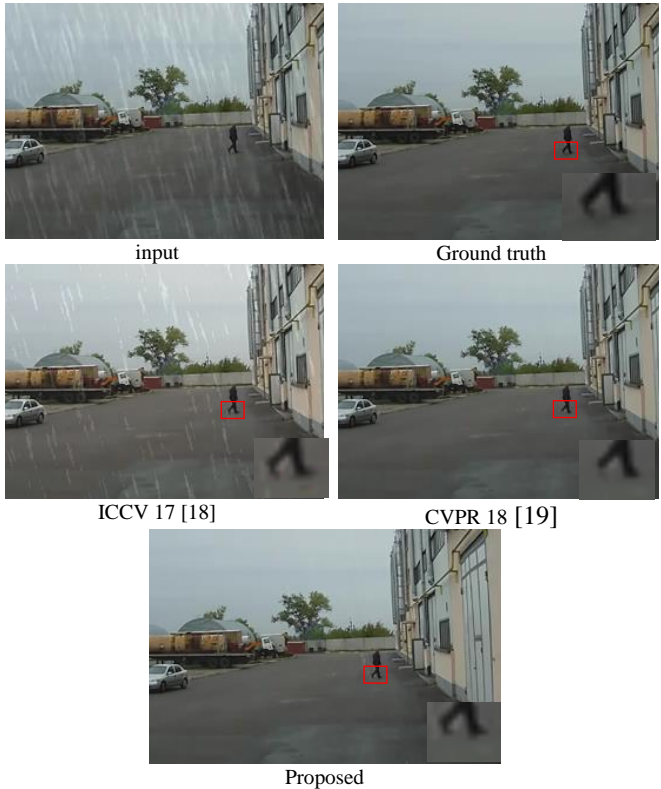

Fig. 14. Rain removal results and comparison between the Proposed method and other methods of video sequence "Park".

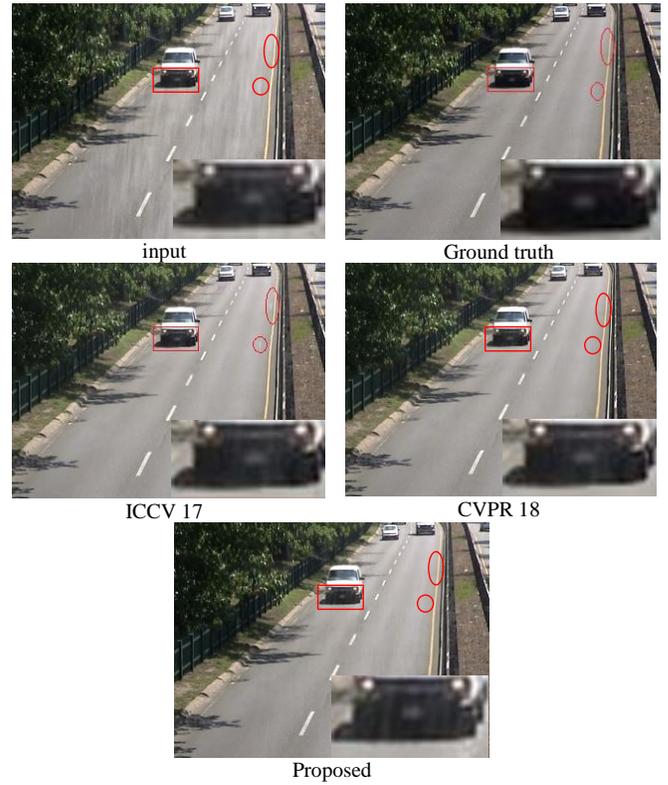

Fig. 15. Rain removal results and comparison between the Proposed method and other methods of video sequence "highway".

Fig. 14 shows the results of another video sequence called "Park" at frame 124. This video has with synthetic rain. The visual result shows the proposed method and CVPR 18 [19] performs very well in rain removal, where ICCV 17 [18] is not as good as the proposed method. The identified areas using red rectangles show that some distortions have been occurred in CVPR 18 [19] method's results. This portion is a part of a moving man's leg. Moreover, the proposed method successfully removes more rain streaks compared to other methods by keeping moving regions quality better.

Fig. 15 shows the results of another video sequence called "highway" at frame 97. This video has with synthetic rain. The visual result shows the proposed method and CVPR 18 performs very well in rain removal, where ICCV 17 is not as good as the proposed method. The results of the proposed method are cleaner.

Figure 16 shows the quantitative comparison of the proposed method against the other two methods using the video sequence called "truck" in terms of PSNR value in each frame. In the figure, the input PSNRs mean the PSNRs of the original frames against ground truth frames (i.e. without rain) which should be the lowest as they have rain. The proposed method outperforms the method in [18] for all frames. However, the proposed method outperforms the method in [19] in most of the frames. These demonstrate that the proposed method not only successfully removes rain from the frames but also keeps the moving object in better quality.

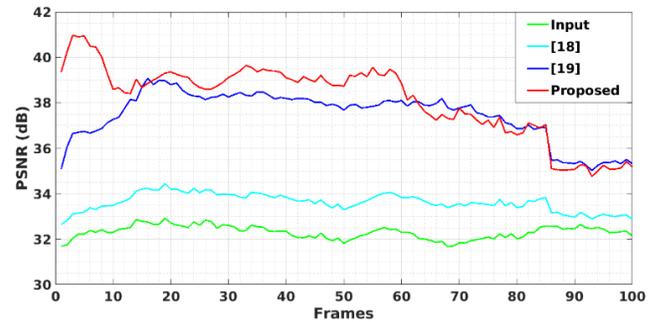

Fig. 16 Quantitative comparison of the proposed method with other relevant methods in terms of frame level PSNR value using truck video sequence.

Table 1 represents the comparison of average PSNR value of all frames for two synthetic datasets "Truck" and "Park". For both of the datasets, the proposed method performs better than other methods.

Table 1 Average PSNR comparison between different methods

| Dataset | Input | ICCV 17[18] | CVPR 18[19] | Proposed |
|---|---|---|---|---|
| Truck | 32.3063 | 33.6189 | 37.4504 | 38.1368 |
| Park | 31.4701 | 25.9783 | 36.1530 | 36.2526 |

## IV. CONCLUSION

In this paper, we try to understand the insightful characteristics of the rain streak and then use them to make a rain-free video.

For these, we identify three crucial characteristics: temporal duration appearance, width, and relative location of the rain streak. The temporal duration appearance is an important phenomenon of a rain streak as the rain streak lasts for a pixel location for a short time, i.e. for a few numbers of frames. The rain streak has a certain width for a low or medium density rain. Moreover, the location of the rain streak is naturally scattered or isolated. We gradually exploit these features to identify rain streak from normal moving regions after separating the rain and moving regions using dynamic background modelling. We also process the features in such a way that they can be applicable for different resolutions and frames of the video sequences. Moreover, we also extract the features and use them in such a way that the proposed method can work in the real-time. To verify the superiority of the proposed method, we use video sequences with both real and synthetic rains and compare the performance against two contemporary and relevant methods. The experimental results confirm that the proposed method outperforms those methods in terms of better rain-free video and quality moving regions.